\documentclass[sigconf,nonacm]{acmart}
\newcommand{\paperTitle}{EDNA-Covid: A Large-Scale Covid-19 Tweets Dataset Collected with the EDNA Streaming Toolkit}

\newcommand{\paperAuthors}{Abhijit Suprem, Calton Pu}

\usepackage{etoolbox}
\usepackage{pifont}
\hypersetup{citecolor=blue,linkcolor=blue}
\usepackage{amsmath,amsopn}
\usepackage{subcaption}
\usepackage{endnotes,microtype,xspace,graphicx,fancyvrb,multirow}
\usepackage{booktabs}
\usepackage{array,underscore,relsize}
\usepackage[T1]{fontenc}
\usepackage{times}
\usepackage{comment}
\usepackage{enumitem}
\usepackage{nicefrac}
\usepackage[labelfont=bf,font=small,skip=5pt]{caption}

\usepackage{xstring}

\usepackage{afterpage}
\usepackage{graphicx, balance, epsfig, endnotes}
\usepackage{colortbl}
\usepackage{tabularx}
\usepackage{listings}

\usepackage[ruled,vlined,linesnumbered]{algorithm2e}
\usepackage[capitalize,noabbrev,nameinlink]{cleveref}

\usepackage{siunitx}
\usepackage{hhline}

\usepackage{tikz}

\newcommand*\circled[2][1.6]{\tikz[baseline=(char.base)]{
    \node[shape=circle, draw, inner sep=1pt, 
        minimum height={\f@size*#1},] (char) {\vphantom{WAH1g}#2};}}
\makeatother

\newcommand{\squishitemize}{
 \begin{list}{$\bullet$}
  { \setlength{\itemsep}{0pt}
     \setlength{\parsep}{3pt}
     \setlength{\topsep}{3pt}
     \setlength{\partopsep}{0pt}
     \setlength{\leftmargin}{1.95em}
     \setlength{\labelwidth}{1.5em}
     \setlength{\labelsep}{0.5em} } }

\newcounter{Lcount}
\newcommand{\squishlist}{
    \begin{list}{\arabic{Lcount}. }
   { \usecounter{Lcount}
        \setlength{\itemsep}{0pt}
        \setlength{\parsep}{3pt}
        \setlength{\topsep}{3pt}
        \setlength{\partopsep}{0pt}
        \setlength{\leftmargin}{2em}
        \setlength{\labelwidth}{1.5em}
        \setlength{\labelsep}{0.5em} } }

\newcommand{\squishend}{\end{list}}

\newcommand{\PP}[1]{
\vspace{2px}
\noindent{\bf \IfEndWith{#1}{.}{#1}{#1.}}
}
\newcommand{\PPP}[1]{
\vspace{2px}
\indent{\it \IfEndWith{#1}{.}{#1}{#1.}}
}

\crefname{algocf}{alg.}{algs.}
\Crefname{algocf}{Algorithm}{Algorithms}

\newcommand{\dataset}{\textsc{EDNA-Covid}\xspace}
\newcommand{\edna}{EDNA\xspace}
\newcommand{\ednajob}{EDNA Job\xspace}
\newcommand{\ednaapp}{EDNA Application\xspace}

\begin{document}

\title{\paperTitle}

\author{\paperAuthors}
\affiliation{%
  \institution{Georgia Institute of Technology}
  \department{School of Computer Science}
  \city{Atlanta}
  \state{GA}
  \postcode{30318}
  \country{USA}
}
\email{{abhijit.suprem, calton.pu}@cc.gatech.edu}

\renewcommand{\shortauthors}{A. Suprem et al.}

\begin{abstract}

The Covid-19 pandemic has fundamentally altered many facets of our lives. 
With nationwide lockdowns and stay-at-home advisories, conversations about the 
pandemic have naturally moved to social networks, e.g. Twitter. 
This affords an unprecedented insight into the evolution of social discourse in 
the presence of a long-running destabilizing factor such as a pandemic with 
the high-volume, high-velocity, high-noise Covid-19 Twitter feed. 
However, real-time information extraction from such a data stream 
requires a fault-tolerant streaming infrastructure to perform the 
non-trivial integration of heterogenous data sources from news 
organizations, social feeds, and authoritative medical organizations 
like the CDC. 
To address this, we present 
(i) the EDNA streaming toolkit for consuming and processing streaming data, and 
(ii) EDNA-Covid, a multilingual, large-scale dataset of COVID-19 
tweets collected with EDNA since January 25, 2020. 
EDNA-Covid includes, at time of this publication, over 600M 
tweets from around the world in over 10 languages. 
We release both the EDNA toolkit and the EDNA-Covid dataset 
to the public so that they can be used to extract valuable 
insights on this extraordinary social event.

\end{abstract}

\begin{comment}
\begin{CCSXML}
<ccs2012>
<concept>
<concept_id>10002951.10003227.10003392</concept_id>
<concept_desc>Information systems~Digital libraries and archives</concept_desc>
<concept_significance>500</concept_significance>
</concept>
<concept>
<concept_id>10002951.10003260</concept_id>
<concept_desc>Information systems~World Wide Web</concept_desc>
<concept_significance>500</concept_significance>
</concept>
</ccs2012>
\end{CCSXML}
\ccsdesc[500]{Information systems~Digital libraries and archives}
\ccsdesc[500]{Information systems~World Wide Web}

% Comment the above block out to hide CCS Concepts or update as per https://dl.acm.org/ccs/ccs.cfm

\keywords{Memento, Web Archiving, Frogs}
\end{comment}

\maketitle

\section{Introduction}
\label{sec:introduction}

Covid-19, also known as the 2019 Novel Coronavirus disease, 
is caused by the SARS-CoV-2 virus. 
It is a rapidly spreading disease that was first reported in Wuhan,
China, in December 2019, and has since spread to all continents 
across the globe. 
It was first picked up as a viral pneumonia in the Wuhan region 
on December 31, 2019. 
WHO signaled the agency's highest level of alarm on January 
30, 2020, by labeling the outbreak as a Public Health Emergency 
of International Concern, and subsequently declared it a 
pandemic on March 11, 2020 after reports of 118K cases 
in 114 countries and a 13x increase in cases outside China within 2 weeks.

Since then, over 33M people have contracted the disease, 
with over 1M fatalities and 24M recoveries, as of September 30, 2020. 
The response to the pandemic has run the gamut, from complete 
nationwide lockdown with strict enforcement, as in the case of China and 
several Western European nations like Italy, France, and Spain, 
to stay-at-home advisories with decentralized enforcement in the 
United States, to no lockdown combined with extensive testing and 
contact tracing, in South Korea, to no lockdown with no federal contact tracing . 

In conjunction, both WHO and national CDCs have both recommended 
a slew of guidelines to slow down the spread and flatten the curve 
to ensure the healthcare industry is not strained with surge in infections.
These guidelines, which include public mask mandates, social distancing, 
work-from-home, cancellation of public events, and shutdown of schools, 
have naturally led to increasing online participation as people 
turn to social media to carry out the conversation \cite{socialincrease}.

This sustained increase in online engagement in reference to a single 
event provides an unprecedented insight into a slew of areas in 
natural language processing, such as social communication modeling, 
credibility analysis, topic modeling, and fake news detection. 
Our \dataset dataset, which contains over 600M tweets from over 10 languages, 
would be an excellent source for research into the social and language 
dynamics of the pandemic. 
Our dataset demonstrates concept drift (see \autoref{fig:drift} in \autoref{sec:drift}), 
making it ideal for testing streaming models of analytics.

Data exhibits concept drift when its underlying distribution changes over time, 
usually over several years. 
Under concept drift, machine learning models and conventional offline 
analytics will degrade as their prediction data desynchronized from 
their training data model. 
Concept drift is a natural part of real data; several examples of drift 
abound in nature, from changing seasons \cite{ebka}, 
which can degrade performance of computer vision systems, 
to lexical drift \cite{lexical}, which can degrade performance of NLP 
models over different geographical regions. 
An important requirement inn concept drift research is data 
that exhibits such drift to enable development and testing 
of drift detection and adaptation mechanisms.

With \dataset, we present a dataset that exhibits concept drift. 
The online discourse on the Covid-19 pandemic has taken root 
in a dizzying array of online communities, such as 
sports \cite{sports}, academia \cite{academic}, and politics \cite{politics}. 
This allows us a firsthand look at a real-world example of concept 
drift as the online conversations change over time to accommodate new 
actors, knowledge, and communities. 
This yields a high-volume, high-velocity data stream with noise and 
drift as the underlying conversations about the pandemic transition 
from confusion to information to misinformation \cite{infodemic} and today, 
with the US election nearing, disinformation \cite{disinformation}.

We will first present \edna, our toolkit for consuming and processing streaming data. 
Then we present \dataset dataset, the streaming methods we employed, and some salient statistics about the dataset.
\section{EDNA}
\label{sec:edna}

\edna is an end-to-end streaming toolkit for ingesting, 
processing, and emitting streaming data. 
\edna is based on our prior work with LITMUS \cite{litmus} and 
ASSED \cite{assed}, and incorporates a slew of improvements for 
faster deployment, fault tolerance, and end-to-end management.
\edna{'s} initial use was a test-bed for studying concept drift 
detection and recovery. 
Over time, it has grown to a toolkit for stream analytics. 
We are continuing to work on it to mature it for production clusters. 
We have released an alpha version at \url{https://github.com/asuprem/edna}. 
In this section, we describe some high-level details about the toolkit. 

\subsection{Architecture}
\label{sec:architecture}

\begin{figure}[t] 
	\centering 
	\includegraphics[width=\linewidth]{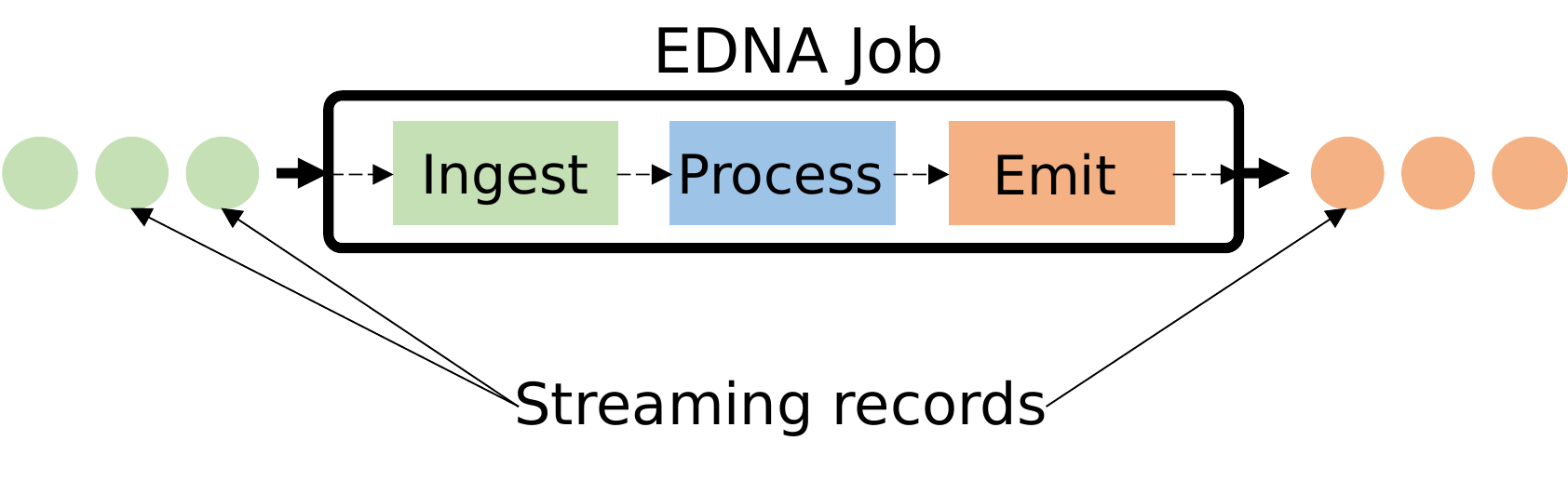}
    \caption{\textbf{An EDNA Job}}
    \label{fig:ednajob}
\end{figure}

The central abstraction in \edna is the \textit{ingest-process-emit} loop, 
implemented in an \ednajob. 
We show an \ednajob in ~\autoref{fig:ednajob}. 
\PP{EDNA Job} 
Each component of the loop in an \ednajob is an abstract primitive in \edna 
that is extended to create powerful operators. 
\squishlist
\item \textbf{Ingest primitives} consume streaming records. 
\item \textbf{Process primitives} implement common streaming transformations 
such as map and filter \cite{flink}. Multiple process primitives can be chained in the same job. 
\item \textbf{Emit primitives} generate an output stream that can be sent to a storage sink, 
such as a SQL table, or to another \ednajob.
\squishend

\begin{figure}[t] 
	\centering 
	\includegraphics[width=\linewidth]{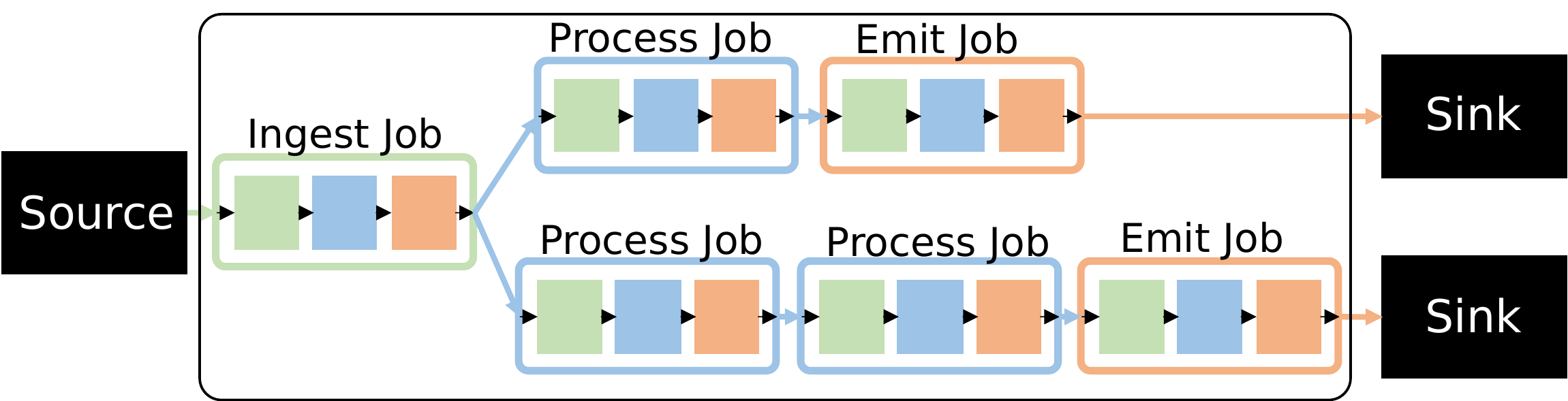}
    \caption{\textbf{An EDNA Application}}
    \label{fig:ednaapp}
\end{figure}

\PP{EDNA Application}
An \ednaapp consists of several jobs in a DAG, as in ~\autoref{fig:ednaapp}. 
We apply the \ednajob abstraction to the application as well, 
where an \ednaapp consists of: 
\squishlist
\item At least 1 Ingest-Job to ingest a stream from external sources, such as the Twitter Streaming API. 
The Ingest-Job should not do any processing to reduce backpressure and ensure the highest throughput for consuming an external stream.
\item Process-Jobs that process the stream. Each Process-Job runs its own \textit{ingest-process-emit} loop to transform the stream
\item At least 1 Emit-Job to emit a stream to external sinks, such as a SQL table, S3 bucket, or distributed file system such as Hadoop.
\squishend

\subsection{The EDNA Stack}
\label{sec:stack}
\begin{figure}[t] 
	\centering 
	\includegraphics[width=\linewidth]{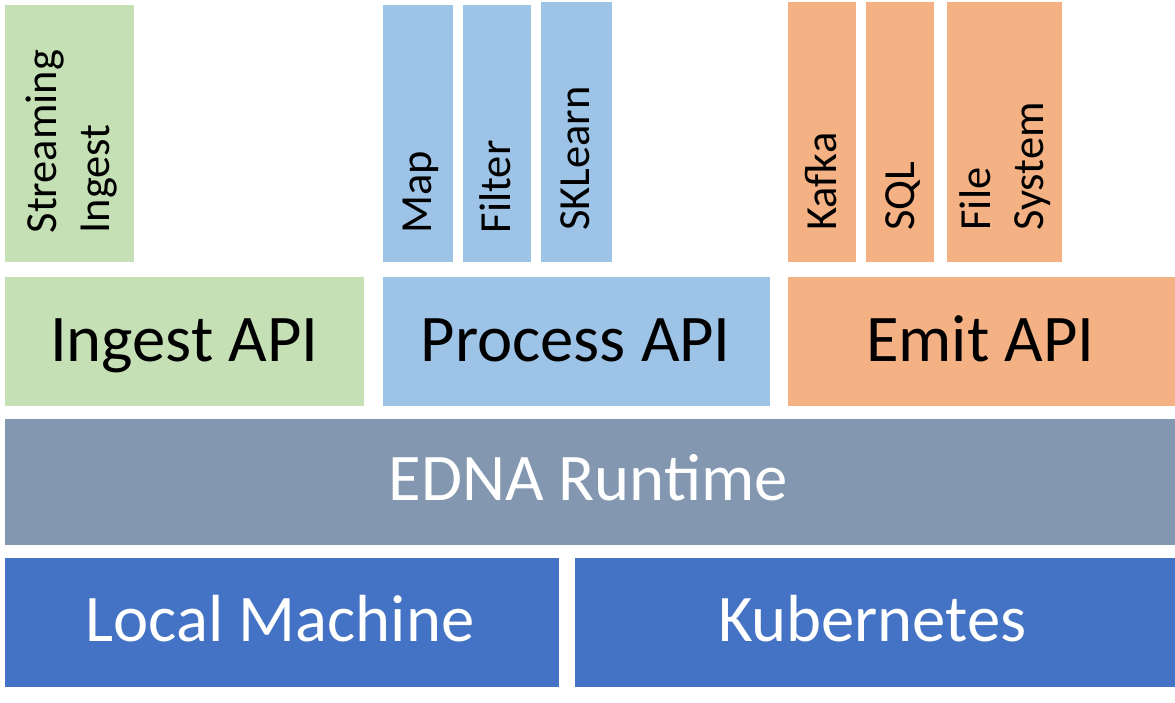}
    \caption{\textbf{EDNA Stack}}
    \label{fig:ednastack}
\end{figure}

The \edna stack (~\autoref{fig:ednastack}) consists of four layers: deployment, runtime, APIs, and plugins.

\edna can be deployed on a local machine for single jobs or on clusters managed by orchestrators 
like Kubernetes for multiple jobs in a streaming application. 
On a cluster deployment, \edna uses Apache Kafka \cite{kafka}, a durable message broker 
with built-in stream playback to connect jobs, and Redis \cite{redis} to share information 
between jobs. 
The \edna runtime manages and executes jobs on the applied deployment. 
\ednajob{s} use the ingest, process, and emit APIs to implement the 
\textit{ingest-process-emit} loop, with the appropriate plugin for complete the job.

The next section describes our \dataset dataset and the \ednaapp we use to generate the dataset.

\section{EDNA-Covid Dataset}
\label{sec:dataset}

We have collected the \dataset dataset since January 25, 2020 
using Twitter's streaming API. 
Over time, we have also enriched our dataset with other similar datasets, 
such as \cite{chen} and \cite{lopez}. 
\dataset is similar ins cale to \cite{chen}; in addition, we also provide our data download method by releasing \edna. 
We also perform additional data cleaning to remove irrelevant and deleted messages. 
Furthermore, our dataset is an order of magnitude larger than \cite{lopez}, 
with over 60M tweets between January and March 2020, compared to 6M for \cite{lopez}. 

\subsection{EDNA Application}
\label{sec:ednaapp}
\begin{figure*}[t] 
	\centering 
	\includegraphics[width=\linewidth]{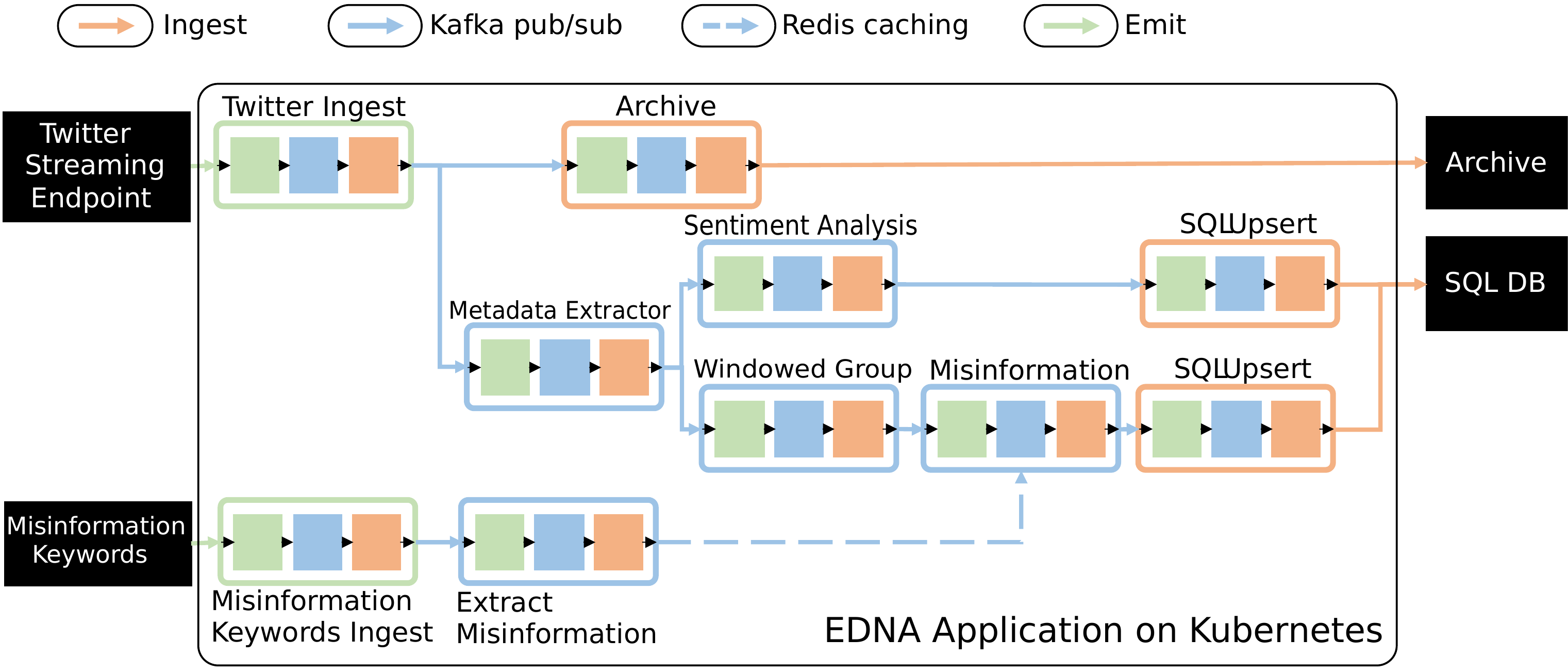}
    \caption{\textbf{EDNA Application for the dataset}}
    \label{fig:ednacovid}
\end{figure*}

We show our \ednaapp to stream tweets for the \dataset dataset in ~\autoref{fig:ednacovid}. 
It consists of the following jobs:
\squishitemize
\item \textbf{Twitter Ingest}: This job connects to the Twitter v2 sampled stream endpoint and 
consumes records for the application. We use the Twitter Sampled Stream API, available at \cite{tweets}. 
This API provides a real-time stream of 1\% of all tweets. 
\item \textbf{Archive}: We immediately archive the raw objects to disk.
\item \textbf{Metadata extractor}: This job extracts the tweet object from the streaming record and performs some 
data cleaning in discarding malformed, empty, or irrelevant tweets. 
Tweets are kept if they contain coronavirus related keywords: 
\textit{coronavirus}, \textit{covid-19}, \textit{ncov-19}, \textit{pandemic}, \textit{mask}, \textit{wuhan}, and \textit{virus}. 
To capture Chinese social data, we also include these keywords in Mandarin. 
We initially included the keyword china during data collection in January and February, 
but decided to omit the phrase since it introduced significant noise, and any 
tweets with the keyword that were relevant to coronavirus already include the above keywords.
\item \textbf{Sentiment analysis}: We use an off-the-shelf tweet sentiment analysis model from 
\cite{sentiment} to record text sentiment. 
We plan to replace this with an EDNA application that will automatically generate and 
retrain a sentiment analysis model with data from Twitter's own streaming sentiment operators.
\item \textbf{SQL Upsert}: This job inserts or updates (if the tweet already exists) 
the tweet object into our database. We record the original fields provided by Twitter, 
plus sentiment and misinformation keywords.
\item \textbf{Windows Group}: We group 1 minute's worth of tweets for faster misinformation 
keyword checking and to record misinformation keyword statistics on a per-minute window.
\item \textbf{Misinformation}: We check whether the grouped tweet objects contain any 
of the misinformation keywords extracted by the \textbf{Extract Misinformation} job (see below). 
The job regularly updates its cache of keywords from Redis.
\item \textbf{Misinformation Keywords Ingest}: We obtain a collection of misinformation keywords from Wikipedia \cite{wikipedia} and from \cite{covidmisinfo}; 
the keywords from the former are obtained with a Wikipedia ingest plugin that reads the misinformation article each day. 
The keywords from the latter are provided directly to the job since they are not updated and do not need to be retrieved repeatedly.
\item \textbf{Extract Misinformation}: This job parses the misinformation article from \textbf{Misinformation Keyword Ingest} 
and extracts keywords from headlines in the \textit{Conspiracy} section. All keywords are updated in the Redis cache.
\squishend

\subsection{Dataset description}
\label{sec:datadescription}
Even with 1\% of the Twitter stream, we are able to collect a 
large-scale dataset of tweets. 
We show in ~\autoref{tab:counts} the tweets collected since January. 
We converted to parallelized Metadata Extractor jobs near the end of 
June to improve our data collection and reduce instances of dropped tweets. 
We also updated our keyword filtering approach to keep tweets that are retweets of matching tweets.

\begin{table}[h]
    \caption{Per-month tweet counts for EDNA-Covid}
    \label{tab:counts}
    \begin{tabular}{lr}
    \hline
    \textbf{Month} & \textbf{No. Tweets} \\ \hline
    2020-01        & 8,714,684           \\
    2020-02        & 25,553,003          \\
    2020-03        & 31,564,785          \\
    2020-04        & 25,498,020          \\
    2020-05        & 26,895,960          \\
    2020-06        & 99,415,221          \\
    2020-07        & 112,215,578         \\
    2020-08        & 113,543,567         \\
    2020-09        & 103,454,256         \\ \hline
\end{tabular}
\end{table}

Our data is skewed towards English language tweets, as we show in ~\autoref{tab:language} 
with the top 5 language categories. 
We also included Chinese and Japanese tweets with keywords in the corresponding languages; 
including Chinese keywords nets us $\sim$25K tweets per month, 
which is less than 0.1\% of the collected tweets, and including Japanese keywords 
adds $\sim$50K tweets per month. This includes enrichment with tweets from \cite{chen,lopez}.

\begin{table}[]
    \caption{Top 5 languages in EDNA-Covid dataset}
    \label{tab:language}
    \begin{tabular}{lrr}
    \hline
    \textbf{Language}                                                                 & \textbf{No. Tweets} & \textbf{Pct Total} \\ \hline
    English                                                                           & 395,109,343         & 63.4\%             \\
    Spanish                                                                           & 76,653,705          & 12.3\%             \\
    \begin{tabular}[c]{@{}l@{}}South Asian \\ (Indonesian, Javan, Malay)\end{tabular} & 23,681,633          & 3.8\%              \\
    French                                                                            & 21,812,030          & 3.5\%              \\
    Portuguese                                                                        & 19,942,427          & 3.2\%              \\ \hline
    \end{tabular}
    \end{table}

\subsection{Concept Drift in EDNA-Covid}
\label{sec:drift}

\begin{figure}[t] 
	\centering 
	\includegraphics[width=\linewidth]{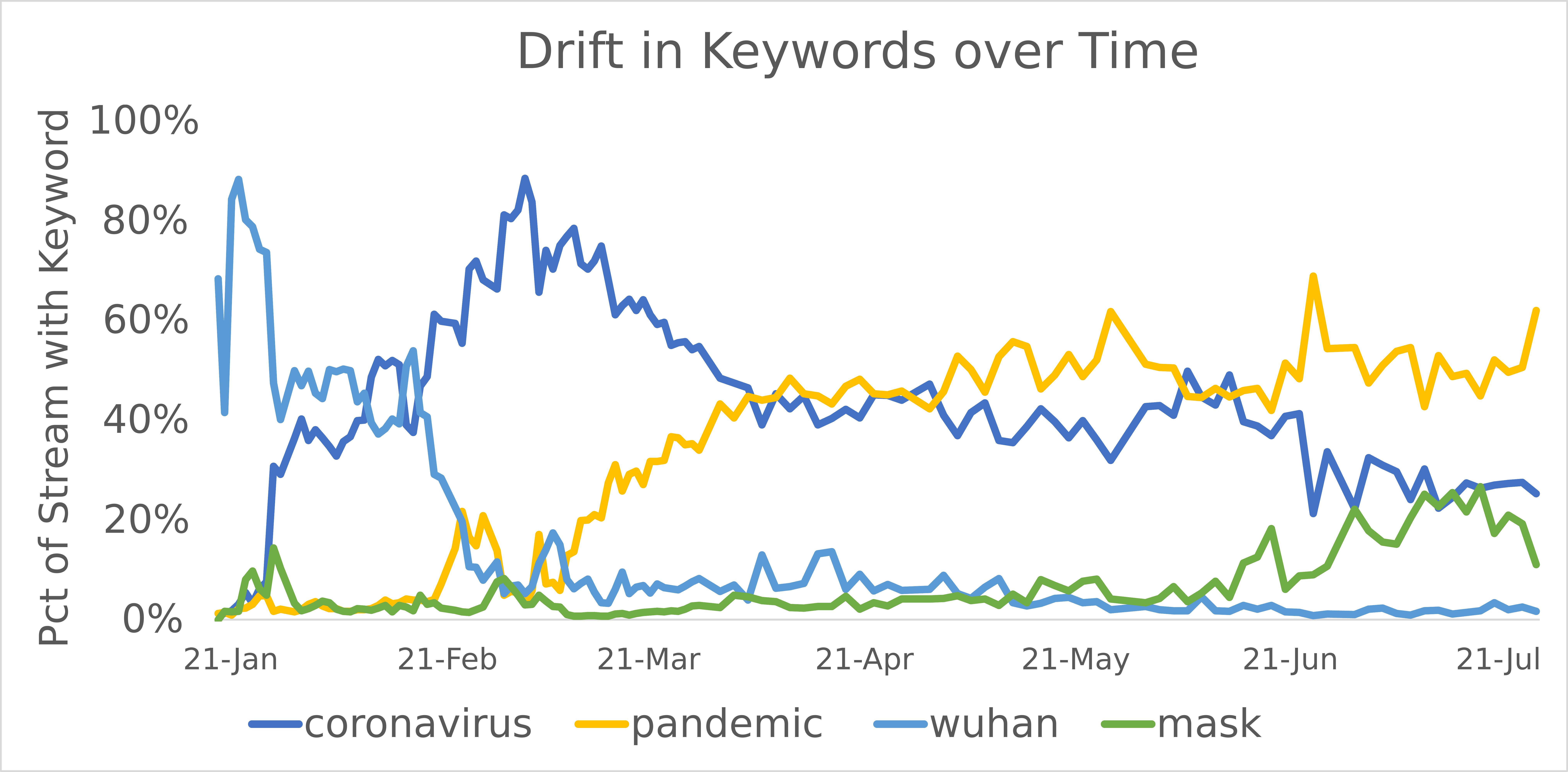}
    \caption{\textbf{Concept Drift in EDNA-Covid}}
    \label{fig:drift}
\end{figure}

Since \dataset is a real-time, multilingual stream of a current event, 
it is ideal for studying concept drift. 
We show an example of drift in ~\autoref{fig:drift}, 
which displays the fraction of the stream that matches our keywords. 
Initially, \textit{wuhan} is a strong keyword for the stream, since the origin 
was an important topic of discussion. 
\textit{Coronavirus} plus related keywords (\textit{covid}, \textit{ncov}) grew in 
importance from the end of January, when WHO declared a global health emergency. 
A sharp increase follows from February 11, when 
WHO formally names the disease caused by the virus as Covid-19. 
\textit{Pandemic} enters the stream on February 21, when WHO reported 
it was preparing for the eventuality, and begins climbing on March 11, 
the date of the pandemic declaration.

In addition, conversations about \textit{wuhan} peak in the first few weeks and then decline, 
signifying the shift in conversations towards the disease itself. 
More recently, conversations about the \textit{pandemic} itself have taken precedence over 
conversations about the disease, as \textit{pandemic} and \textit{mask} see increased weight in the stream. 
As masks have become a more contentious issue, there has been a resurgence 
in conversations about masks, necessitating adjustments to data collection, 
cleaning, and misinformation detection starting July, 2020.

\subsection{Dataset release}
Due to Twitter TOS regarding release of tweets, we are releasing only the Tweet IDs of the 
dataset to the public through a registration method. 
We have provided a form at \url{https://forms.gle/dFYhuMzyPMunY17H9} for dataset requests. 
We will then provide Tweet IDs of our collected tweets. 
Tweets need to be hydrated first using tools like twarc \cite{twarc}. 
We have also provided a sample of TweetIDs for the first few months at \url{https://github.com/asuprem/EDNA-Covid-Tweets}.

\section{Acknowledgements}
This research has been partially funded by National Science Foundation by CISE/CNS (1550379, 2026945, 2039653), SaTC (1564097), SBE/HNDS (2024320) programs, and gifts, grants, or contracts from Fujitsu, HP, and Georgia Tech Foundation through the John P. Imlay, Jr. Chair endowment. Any opinions, findings, and conclusions or recommendations expressed in this material are those of the author(s) and do not necessarily reflect the views of the National Science Foundation or other funding agencies and companies mentioned above.

\bibliographystyle{ACM-Reference-Format}
\bibliography{main} 

\end{document}